\documentclass[11pt,twoside]{article}
\usepackage{asp2006}
\usepackage{epsf}
\usepackage{graphics}
\def\gs{\mathrel{\raise0.35ex\hbox{$\scriptstyle >$}\kern-0.6em
\lower0.40ex\hbox{{$\scriptstyle \sim$}}}}
\def\ls{\mathrel{\raise0.35ex\hbox{$\scriptstyle <$}\kern-0.6em
\lower0.40ex\hbox{{$\scriptstyle \sim$}}}}

\newcommand{\um}{\,$\mu$m}

\newcommand{\lsun}{\,$\rm{L}_{\odot}$}

\def\edcomment#1{\iffalse\marginpar{\raggedright\sl#1\/}\else\relax\fi}
\reversemarginpar

\begin{document}
\title{Detection of the 3.3\um\ PAH feature as well as water ice and HAC absorption in $z$\,$\sim$\,2 ULIRGs}
\author{A. Sajina, L. Yan, H. Spoon, D. Fadda}
\affil{Haverford College, Haverford, PA, 19041}
\affil{WISE, California Institute of Technology, Pasadena, CA 91125}
\affil{Cornell University, Astronomy Department, Ithaca, NY, 14853}
\affil{NASA {\sl Herschel} Science Center, California Institute of Technology, Pasadena, CA 91125}

\begin{abstract}

We present preliminary results from the highest available signal-to-noise rest-frame 2-8\um\ spectra of $z$\,$\sim$\,2 ULIRGs. Our 10 targets are selected for their deep silicate absorption features based on previous shallower IRS spectra.  The goal of this follow-up program is: 1) allow for a more accurate analysis of inner/hot dust continuum, 2)  detecting the 3.3\um\ and 6.2\um\ PAH features, and 3) detecting molecular absorption features such as due to water ice and hydrocarbons (HACs).  We find that the 3.4\um\ HAC absorption feature is observed in four sources, while the 3.05\um\ water ice feature is observed in three of the sources. The HAC detectability is higher and ice detectability lower than expected from local ULIRGs, but consistent with a more AGN-dominated sample such as this one.  Where ice is detected, the ice-to-silicate ratio is somewhat lower than many local ULIRGs implying on average thinner ice mantles. One source shows the highest redshift reported detection of the 3.3\um\  PAH feature (along with a previously detected 6.2\um\ feature) whose strength is as expected for a starburst-dominated ULIRG.  

\end{abstract}

\vspace{-0.5cm}
\section{Introduction}

$\indent$One of the most exciting recent discoveries is the strong evolution of both the star-formation rate density, and black hole growth from $z$\,$\sim$\,1\,--\,3 to today \citep[for a recent review see][]{soifer08}.  A significant fraction of high-$z$ star-formation and black hole accretion probably takes place in ULIRGs.  But are these high-$z$ ULIRGs scaled-up versions of local analogues, or is this prodigious activity driven by different physical conditions? In interpreting the observations we see at these redshifts, we hope that comparison with local templates is valid. There is an inherent assumption here that the dust properties and composition in these $z$\,$\sim$\,2 sources are consistent with those observed nearby.  One rough test of this assumption is to see whether or not the detectability and relative depth of absorption features arising from different dust components are consistent with those of our local analogs.  The mid-IR regime is well suited to such a test as it contains a wealth of molecular absorption features. By far the strongest is the 9.7\um\ silicate absorption feature. The large silicate grains, are thought to be frequently coated in various ices including water. The water ice can be seen in the 3.05\um\ and 6.15\um\ features. The other major dust component are the small carbon particles. In extreme environments such as observed toward the Galactic Center, the degree of hydrogenation is such that prominent hydrocarbon (HAC) features the strongest being at 3.4\um) are observed \citep[e.g.][]{pendleton94,whittet03}.  Studies of these features in external galaxies include the {\sl ISO} study of dusty galaxies by \citet{spoon02}, the Subaru study of the 3\,--\,4\um\ spectra of 37 ULIRGs \citep{imanishi06} supplemented by the $\sim$\,5\,--\,30\um\ {\sl Spitzer} IRS study of local ULIRGs  \citet{imanishi_irs}, and finally the VLT study of the 3\,--\,4\um\ spectra of 11 more ULIRGs by \citet{risaliti06}.  More recently,  AKARI $\sim$\,2\,--\,5\um\ spectra of 47 ULIRGs was released \citep{imanishi08}. We draw on all these samples for our local comparisons. 

Here, we present the first detections of such molecular absorption features as well as the 3.3\um\ PAH emission feature from a sample of 10 $z$\,$\sim$\,2 ULIRGs with deep {\sl Spitzer} IRS SL1 and LL2 (rest-frame $\sim$\,2\,--\,8\um) spectra. We compare the results with studies of local ULIRGs. 

\section{Our sample of $z$\,$\sim$\,2 sources with IRS spectra}

The sources discussed here come from two separate {\sl Spitzer} Infrared Spectrograph \citep[IRS;][]{houck04} samples, which we call 'GO1' and 'GO2' based on the {\sl Spitzer} epoch of the observations. The G01 sample is the `bright and red` sample. The sample selection and IRS data are discussed in more detail in \citet{yan07}. The key points are: it consists of  52 sources selected in the $\sim$\,3.5\,sq.degrees {\sl Spitzer} Extragalactic First Look Survey (xFLS) using the criteria $F_{24}$\,$>$\,0.9\,mJy,  $\nu F_{\nu}$(24)/ $\nu F_{\nu}$(8)\,$>$\,0.5, $\nu F_{\nu}$(24)/ $\nu F_{\nu}$(0.64)\,$>$\,1.  The bulk of this sample (48/52) has redshifts based on either rest-frame optical Keck/Gemini spectra or the IRS spectra themselves.  Roughly a third of the sample has redshifts of $z$\,$\sim$\,1 and two thirds have $z$\,$\sim$\,2.  

The GO2 sample, is  the `flux-limitedÕ sample. The selection, IRS data, and redshift distribution are discussed in Dasyra et al. (in prep.). The key points are: it consists of 150 sources selected again in the xFLS but largely based on $F_{24}$\,$>$\,0.9\,mJy.  In addition, optically-bright sources were excluded ensuring that $z$\,$\sim$\,0 sources were not included (this also excludes type 1 AGN). This sample is dominated by $z$\,$\sim$\,0.5 sources with a tail extending to $z$\,$\sim$\,2. 

Since the GO2 sample excludes the bright and red sources that were observed in the GO1 sample, the combination of the two results in a more truly flux-limited and color-unbiased sample (Yan et al. 2009, in prep.).  In total, we have $\sim$\,70 IRS sources at 1.5\,$<$\,$z$\,$<$\,2.8 with $L_{\rm{IR}}$\,$\sim$\,$10^{12.4}$\,--\,$10^{13}$\lsun.  These represent an exceptional opportunity to study the  mid-IR properties of $z$\,$\sim$\,2 sources at the high-end of the luminosity function. Their IRS spectra show a range of spectral types dominated by strong-PAH, strong 9.7\um\ silicate absorption, or a power-law continuum \citep[see][]{sajina07a}. The later two types make up the bulk of the sample and are generally believed to signify AGN-dominated sources. Our mid-IR spectral classification is reinforced by an analysis of their rest-frame optical spectra as well as their long-wavelength SEDs \citep{sajina08}. By virtue of their luminosities ($L_{\rm{IR}}$\,$>$\,$10^{12}$\lsun), as well as relative optical faintness, and lack of broad lines in the available optical spectra, they are consistent with being obscured quasars. The strong-PAH subsample overlaps the sub-mm galaxies population (based on their mm-wave fluxes, see Sajina et al. 2008) and therefore this sample includes both the likely significant $z$\,$\sim$\,2 obscured quasars population and starburst-dominated sub-mm galaxies as well as transition objects.  

Out of this combined sample of 70 $z$\,$\sim$\,2 sources with IRS spectra, 10 sources were selected for a {\sl Spitzer} GO4 follow-up with the IRS SL1 and LL2 modules (rest-frame $z$\,$\sim$\,2\,--\,8\um) to reach the depth and wavelength coverage necessary for the observation of the molecular absorption features discussed above ($\sim$\,0.5\,--\,2.5\,hours per source per module depending on source brightness). The sources were selected to have significant silicate absorption features ($\tau_{9.7}$\,$>$\,1) but otherwise to cover a range in SED shapes as sampled by the IRAC8\um\ flux ($\sim$\,3\um\ rest-frame). Most of the sources have weak or no PAH emission, but one strong-PAH source (MIPS22530) was also included (see below). The full analysis of this data is still a work in progress (Sajina et al. 2009 in prep.). Below we show preliminary results showing the detection of the molecular absorption features expected as well as evidence for the 3.3\um\ PAH emission -- the highest-redshift such observation as far as we are aware. 
 
\begin{figure}[!ht]
\begin{center}
\plottwo{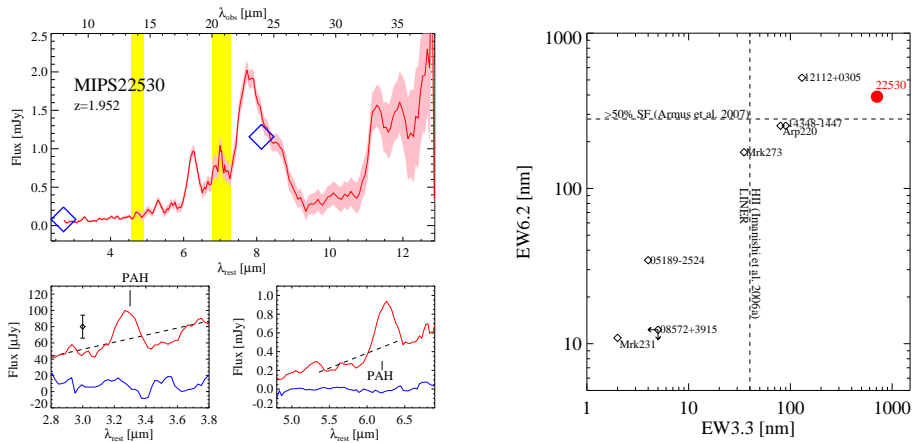}{pah_ratio.eps}
\end{center}
\caption{{\it Left:} The mid-IR spectrum of MIPS22530 ($z_{\rm{spec}}$\,=\,1.952). The pink shaded area shows the 1\,$\sigma$ spread. The yellow rectangular bands show the areas where the different IRS orders are stitched.  The spectrum inside these is not reliable. The panels below are zoom-ins to the 3.3\um\ and 6.2\um\ PAH features. The blue spectra underneath are the background `sky' spectra indicating the noise levels and the fact that the source continuum is detected. {\it Right:} A comparison between the PAH equivalent widths of the 3.3 and 6.2\um\ features of MIPS22530 with those of well known local ULIRGs. The dashed lines indicate the 3.3 and 6.2\um\ equivalent width starburst-ULIRG criteria adopted by Imanishi et al. (2006) and Armus et al. (2007) respectively.\label{pah33}}
\end{figure}

\section{A detection of the 3.3\um\ PAH feature at $z$\,$\sim$\,2}

The one strong-PAH source in this sample, MIPS22530 \cite[see][]{sajina07a} shows a strong 3.3\um\ PAH emission feature with an EW of $\sim$\,700\,nm (see Figure\,\ref{pah33}).  This is consistent with a starburst \citep[see][]{imanishi06}. The ratio of the 6.2 to 3.3\um\ PAH features in this source is consistent with that observed in local starburst-dominated ULIRGs. Its total infrared luminosity however is close to $10^{13}$\lsun\ \citep{sajina08} which is nearly an order of magnitude larger its local counterparts and thus this further confirms our earlier conclusion of scaled-up starbursts among our $z$\,$\sim$\,2 sample \citep{sajina07a}. 

\begin{figure}[!ht]
\begin{center}
\plottwo{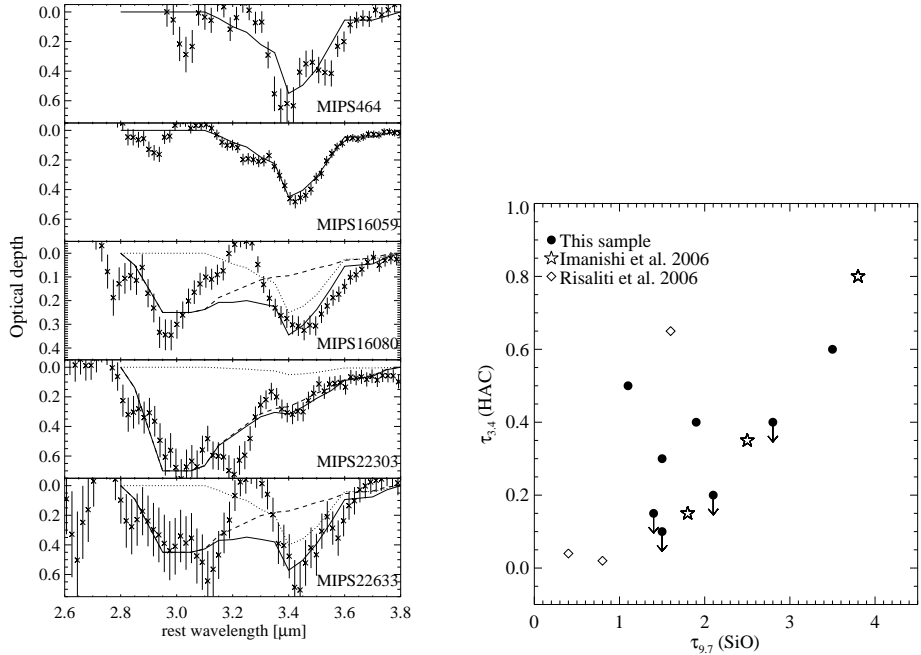}{hac_si.eps}
\end{center}
\caption{{\it Left:} Optical depth profiles in our sample. The dashed and dotted lines are respectively the 3\um\ water ice and 3.4\um\ hydrocarbon absorption profiles from \citet{chiar02}. The excess observed in MIPS16080 and MIPS22633 could be due to 3.3\um\ PAH emission. {\it Right:} A comparison between our $z$\,$\sim$\,2 sources and local ULIRGs suggests that the dust composition is not dramatically different. \label{fig_hac}}
\end{figure}

\section{Molecular absorption features: hydrocarbons and water ice}

Figure\,\ref{fig_hac} shows the optical depth profiles of five of the sources with deeper data where clear water ice and or hydrocarbon absorption features can be seen. The 3.4\um\ feature in particular is exclusively observed in AGN ULIRGs locally \citep{risaliti06}. Most local ULIRGs (whose luminosities are just over $10^{12}$\lsun) are starburst-dominated leading to the low detectability of this feature among them. Figure\,\ref{fig_hac}{\it right} shows that, the ratio of the 3.4\um\ to 9.7\um\ optical depths for our sources is comparable to that of the local ULIRGs with 3.4\um\ detections. The ice-to-silicate ratios on the other hand tend to be on average lower than those of local ULIRGs implying thinner ice mantles. This is consistent with this sample being largely AGN-dominated. Among local ULIRGs, many of the sources with the deepest ice absorption are starburst-dominated sources which are not well represented here. These results suggests that, at least roughly, the general assumption that the dust mixture in distant ULIRGs is the same as observed locally is valid. 

\section{Summary \& Conclusions}

1) We have assembled a substantial ($\sim$\,70 sources) library of IRS spectra of 24\um-bright $z$\,$\sim$\,2 sources ranging between starburst-dominated (SMG-like) to AGN-dominated sources (obscured quasars). In order to better understand the dust properties of the deep silicate absorption sources in particular, we have followed-up 10 such sources to obtain high signal-to-noise spectra in the rest-frame 2\,--\,8\um\ regime. \\
2) In the deeper 2\,--\,8\um\ spectra, we find evidence for water ice, HAC and PAH3.3. Comparisons with local ULIRGs suggest the dust composition for these sources is roughly consistent with the local Universe, considering this sample is largely more AGN-dominated than the typical local ULIRGs. The lack of evolution in the dust composition is an underlying assumption in much of our interpretation of high-$z$ data. \\

\end{document}